\documentstyle[]{article}
\begin{document}
\begin{center}{  \Huge \bf TRANSITION OF THE UNIFORM STATISTICAL FIELD ANYON STATE \\ TO
THE NONUNIFORM \\ ONE AT LOW  PARTICLE \\ DENSITIES.
}
\end{center}
\vskip1cm
\begin{center}
	{\huge   M. Hudak{\footnote{\Large Stierova 23, Kosice} } \\
		\vskip0.5cm
		O. Hudak{\footnote{\Large Department of Aerodynamics and Simulations, Faculty of Aviation, Technical University Kosice, Kosice} } 
	}
\end{center}

\newpage
\section*{Abstract}
Using results of our exact description of the spinless 
fermion motion in a nonhomogeneous magnetic field \( {\bf B}  =
B( 0, 0, 1/cosh^{2}( \frac{x-x_{0}}{ \delta })) \) we study a gas
of these particles moving in this field.
For lower densities \(  \nu <  \nu_{c}(B, \delta ) \) the corresponding
total energy is lower than that of the uniform field state. 
Thus when the density of anyons decreases a transition from the
uniform statistical field state to the nonhomogeneous field 
state is predicted.

\newpage
\section*{}
Non-Abelian anyons and topological quantum computation has recently emerged as one of the most exciting approaches to constructing a fault-tolerant quantum computer \cite{[0]}. Strongly correlated quantum systems can exhibit behavior called topological order which is characterized by non-local correlations that depend on the system topology. Such systems can exhibit 
phenomena such as quasi-particles with anyonic statistics and have been proposed as candidates for naturally fault-tolerant quantum computation. Despite these remarkable properties, anyons have not been observed in the year 2008 directly \cite{[00]}. Recently it was presented an experimental emulation of creating anyonic excitations in a superconducting circuit that consists of four qubits, achieved by dynamically generating the ground and excited states of the toric code model, i.e., four-qubit Greenberger-Horne-Zeilinger states. The anyonic braiding is implemented via single-qubit rotations: a phase shift of π related to braiding, the hallmark of Abelian 1/2 anyons, has been observed through a Ramsey-type interference measurement \cite{[01]}.

Neverthless it is interesting to study spinless 
fermion motion in a nonhomogeneous magnetic field \( {\bf B}  =
B( 0, 0, 1/cosh^{2}( \frac{x-x_{0}}{ \delta })) \)
of these particles moving in this field. 
For lower densities \(  \nu <  \nu_{c}(B, \delta ) \) the corresponding
total energy is lower than that of the uniform field state. 
Thus when the density of anyons decreases a transition from the
uniform statistical field state to the nonhomogeneous field 
state is predicted. When the density of anyons present in this system decreases a transition from the uniform statistical field state to the nonhomogeneous field 
state is predicted. This may forward our observation possibilities to observe anyon properties.

An homogeneous magnetic field  {\bf B} = ( 0, 0, B) strongly
influences states of charged fermions moving in an x-y plane 
perpendicular to the field direction. Landau energy levels 
and their degeneracy characterize this motion, \cite{[1]}. A gas
of free spinless fermions has its total energy \( E_{T}( B, \nu ) \)
larger or equal to its total energy  \(  E_{T}( 0, \nu ) = 2 \pi tN \nu^{2} \)
in the zero field:
\begin{equation}
\label{eq:1}
\Delta E_{h}(n) \equiv E_{T}( B, \nu )-E_{T}(0, \nu ) = 2 \pi tN( \nu_{n+1} -
\nu )( \nu - \nu_{n})
\end{equation}
where \( \nu_{n+1} \geq \nu > \nu_{n},  \nu_{n} \equiv n. \frac{ \Phi}{ \Phi_{0}},
n= 0, 1, ...;  \Phi_{0} \) is a unit of the magnetic flux, \( \Phi \equiv Ba^{2},
t \equiv \frac{ \hbar^{2} }{2ma^{2}}. \) Here \( \nu \) is the number
density of the gas, \( \nu \equiv \frac{N_{f}}{N} \), \( N_{f} \) is the total
number of fermions, \( Na^{2} \) is the area of the square with the side
length \( L=a \sqrt{N}, \) to which the motion is bounded, m is the fermion mass.
Note that \( a \) is a characteristic length of the system, its value is of
the order of a lattice constant value.
The energy level degeneracy occurs only if a number of Landau 
levels is completely filled (e.i. if for some n \( \nu_{n}= \nu ). \) Recently it was
shown in \cite{[2]} that in the presence of a periodic lattice
potential the ground state energy of a gas of spinless fermions
in an uniform magnetic field in the vicinity of the 
filled lowest Landau level is lower than that in zero field.
This problem was studied further in context of commensurate 
flux phases, \cite{[3]}. If a nonhomogeneity of the field is introduced by a
local field intensity decrease then competition of two 
tendencies is expected to occur: a decrease of the single
fermion energy level due to decreased value of the field 
and a decrease of the every energy level degeneracy due to 
larger spacing between centers of neighboring orbits within
the region of smaller fields. Spectrum of 2d Bloch electrons
in a periodic magnetic field was studied in \cite{[BBR]}. Using
semiclassical methods authors of this later paper investigated the case where
the magnetic unit cell is commensurate with the lattice unit cell. Their work
is in some sense extension of previous studies of free electrons in
periodic magnetic field \cite{[N]} to the lattice case. Our aim in this paper is
to  present results of our study of the motion of a spinless
fermion gas bounded to the square LxL in a nonhomogeneous
static magnetic field perpendicular to this plane. We neglect 
the lattice periodic potential influence on the gas energy 
spectrum in this paper. We consider in more details the limit in which
nonhomogeneity disappears and a uniform field appears.
In difference to \cite{[BBR]}, \cite{[N]} and \cite{[6]}
we do not consider a periodic magnetic field.  Recently, \cite{[4]}, an exact
description of motion of the quantum spinless fermion in a nonhomogeneous
magnetic field described by the vector potential \( {\bf A} =
( 0, B \delta tanh( \frac{x-x_{0}}{ \delta}), 0) \) was found.
We use these results in this paper to study the stability of the statistical
uniform anyon state with respect to a nonuniform field state. Firstly using
the single fermion energy from \cite{[4]} we find a total energy of a gas of
spinless fermions moving in our nonhomogeneous field. Then we compare this
energy with the total energy of the same gas moving in the uniform field
with the same intensity B. We have found that at low densities
\( \nu < \nu_{c}(B, \delta) \) the nonhomogeneous field state of the anyon gas
is preferred. Occurence of such a kind of instability has consequences for
interpretation of recent experiments \cite{[5']} searching for the T- and
P- symmetry breaking phenomena due to presence of particles with exotic
statistics - anyons.

        In the case of motion of a quantum
spinless fermion in a nonhomogeneous magnetic field described by the vector
potential \( {\bf A} = ( 0, B \delta tanh( \frac{x-x_{0}}{ \delta}), 0) \)
the energy spectrum of the motion in the x-direction is splitted, see in
\cite{[4]}, into a discrete and a continuous parts for general values of
the field B and of the nonhomogeneity parameter \( \delta. \)
We take \( x_{0}=0 \) in the following, thus
field has its maximum intensity at \( x=0. \) Let us consider
the limit of strong fields \( ( F \equiv 2 \pi \frac{ \Phi^{'}}{ \Phi} >>
1/2, \) where  \( \Phi^{'} \equiv B \delta^{2}) \) in which case
a linear type nonhomogeneity is localized near the two edges
\( x = \pm L/2, \) if \( \delta >> a \) keeping L finite.
In this limit it is sufficient to take into account the lowest energy levels
of the spectrum.
The eigenvalues of the energy corresponding to
this part of the spectrum are given by, see in \cite{[4]} :
\[ E_{n}(p)=
\frac{p_{y}^{2}}{2m}(1- \frac{F^{2}}{((\frac{1}{4}+F^2)^{\frac{1}{2}}-((1/2)+n)
)^{2}})+ \]
\[ ( \frac{ \hbar^{2}}{2m \delta^{2}})(F^{2}-((\frac{1}{4}+F^{2})^{\frac{1}{2}}
-( \frac{1}{2}+n))^{2}, \]
where \( n= 0, 1,...[n_{max}], \) here [n] denotes an integer part of a real
number n, \( p_{y} \) is the y-momentum. Let us define \( P \equiv
\frac{ \mid p_{y} \mid \delta }{ \hbar } \). The number \( n_{max} \) is defined
by:
\[ n_{max} = ( \frac{1}{4}+F^{2})^{ \frac{1}{2} }-(1/2)-( \mid P \mid F)^
{ \frac{1}{2}}, \]
for given values of P and F.

The limit of strong but still nonhomogeneous field
is achieved for \( F \longrightarrow \infty \) keeping the nonhomogeneity
parameter \( \delta \) finite while increasing the field intensity B,
\( B \longrightarrow \infty \).
For \( F^{2} >> \frac{1}{4} \) and for small quantum numbers n
the energy \( E_{n}(p_{y}) \) expanded into series of 1/F powers takes
the form :
\[ E_{n}(p_{y}) \approx \hbar \omega (n+ \frac{1}{2})- \\
( \frac{ \hbar^{2}}{2m \delta^{2}} )
((n+ \frac{1}{2})^{2}+ \frac{1}{4})- \frac{p_{y}^{2}}{mF}(n+ \frac{1}{2})+ \]
\[( \frac{ \hbar^{2}}{8mF \delta^{2}})(n+ \frac{1}{2})+ O( \frac{1}{F^{3}}). \]
where \( \omega \equiv \frac{Bc}{em} \) is the cyclotron frequency.
We see that the energy levels are degenerated in the
limit of strong but modulated fields if the energy expansion above
is restricted to the first two terms, which are of the \( F^{1} \) and \( F^{0} \)
orders respectively.
The largest value of the third term in this expansion is negligible with respect
to the second term
\[  max( \frac{p^{2}_{y}}{mF}(n+ \frac{1}{2})) <<
( \frac{ \hbar^{2}}{2mF \delta^{2}})(n+ \frac{1}{2}). \]
if we take into account that there exists a natural cut-off
for \( p_{y} \) momenta, \( max( \mid p_{y} \mid ) = \frac{ \pi \hbar }{ a}, \)
due to the underlying crystal and if we assume that the field intensity B
satisfies the inequality:
\[ \frac{ \Phi}{ \Phi_{0} } >>  8 \pi^{2}, \]
where \( \Phi \equiv B.a^{2} .\)
If the third term and the following terms are not taken into account
in calculations of the energy \( E_{n}(p_{y}) \) then the degeneracy of
the n-th level appears due to the lost of the energy dependence on \( p_{y} \)
momentum. One can say that these levels are, \cite{[4]}, modified
Landau levels with energies in the form:
\begin{equation}
\label{eq:2}
E_{n} = \hbar \omega (n + \frac{1}{2}) - \frac{ \hbar^{2}}{2m \delta^{2} }
[(n + \frac{1}{2})^{2} + \frac{1}{4}] + O(1/F),
\end{equation}
where  \[ n = 0, 1, ...  <<  n_{m}; n_{m} \approx F. \]
Note that
\[ \frac{ \hbar^{2}}{2m \delta^{2}} = 4t( \frac{L}{2 \delta})^{2}/N. \]
From (\ref{eq:2}) we see that in the strong
nonhomogeneous magnetic fields the neighboring energy levels 
are not equidistant as in the uniform field case. Every 
energy level \( E_{n} \) remains degenerated within considered
approximation, its degeneracy \( D_{n} \) is found to be:
\begin{equation}
\label{eq:3}
D_{n} = D_{L} \frac{tanh( \frac{L}{2 \delta })}{ \frac{L}{2 \delta }},
\end{equation}
if the characteristic length L and the nonhomogeneity parameter \( \delta \)
satisfy
\[ tanh(L/2 \delta ) < (1- \frac{2}{F}(n +  \frac{1}{2})). \]
Here \( D_{L} \equiv \frac{Bea^{2}}{hc} N \) is the Landau level degeneracy
as it is given in the case of the uniform field. The form of the degeneracy
\( D_{n} \) given above holds for all orders of F. However, the large F
expansion in (\ref{eq:2}) limits its validity to the region of system parameters
given by the inequality below (\ref{eq:3}). This inequality follows from
the usual, \cite{[5]}, boundary conditions: periodicity in the y-direction
perpendicular to the x-axis and limits on the position of
the orbit center in the x-direction to the region \( < -L/2, +L/2 >. \)
The orbit center x-coordinate \( x_{c} \) is given, \cite{[4]}, by
\[ tanh(x_{c}/ \delta ) = ( \frac{-p_{y} \delta }{ \hbar})/F. \]
Note that this relation also reflects the fact that closed particle orbits
of their motion in our magnetic field do exist only in the limited
region of system parameters and of the \( p_{y}-momentum \) such that
tanh function above is note larger (or smaller) than 1 (than -1).

          Straightforward calculations of the ground state
energy \( E_{T}(B, \delta , \nu ) \) for spinless fermion gas with density
\( \nu \) in the limit of strong but nonhomogeneous fields specified by
\( B, \delta \) lead to the modification of (\ref{eq:1}). We have found
that the energy difference between the nonhomogeneous field state and
the zero field state:
\[ \Delta  E_{nh}(n) \equiv  E_{T}(B, \delta , \nu ) - E_{T} (0, \nu ) \]
is given by the following expression:
\begin{equation}
\label{eq:4}
\Delta  E_{nh}(n) = 2 \pi t N [( \nu -  \nu_{n}
\frac{tanh( \frac{L}{2 \delta } )}{ \frac{L}{2 \delta } } )
( \nu_{n+1} \frac{ tanh( \frac{L}{2 \delta } ) }{ \frac{L}{2 \delta } } -  \nu ) +
\end{equation}
\[ ( 1 - \frac{tanh( \frac{L}{2 \delta } ) }{ \frac{L}{2 \delta } })( \nu (2 \nu_{n}+
 \nu_{1}) - \frac{tanh( \frac{L}{2 \delta } ) }{ \frac{L}{2 \delta } }
\nu_{n+1} \nu_{n} )] - \]
\[ - \frac{ta^{2}}{ \delta^{2} } N [ \nu (n^{2} + n + \frac{1}{2}) -  \nu_{n}
( \frac{2n^{2}}{3} + n + \frac{1}{3} )]. \]
The total energy difference (\ref{eq:4}) is found assuming that there are
n levels \( 0, 1, ..., n-1 \) filled
and that the n-th level is filled partially. The gas
density \( \nu \) in (\ref{eq:4}) is limited by the following inequalities:
\begin{equation}
\label{eq:4'}
\nu_{n+1} \frac{ tanh( \frac{L}{2 \delta } ) }{ \frac{L}{2 \delta } } \geq \nu >
\nu_{n} \frac{ tanh( \frac{L}{2 \delta } ) }{ \frac{L}{2 \delta } },
\end{equation}
\[ \nu_{n} \equiv n \Phi / \Phi_{0}. \]
The uniform field result (\ref{eq:1})
follows from (\ref{eq:4}) and (\ref{eq:4'}) in the
limit \( \delta \longrightarrow \infty \) keeping values of
all the other system parameters constant.

         When only the lowest energy level \( n=0 \) is filled partially we find
from (\ref{eq:4}) and (\ref{eq:4'}) that:
\begin{equation}
\label{eq:4''}
\Delta E_{nh}(0)= 2 \pi tN \nu ( \nu_{1} - \nu ) - N \nu t
( \frac{a}{ \delta })^{2}/2,
\end{equation}
where
\[ \nu_{1} \frac{tanh( \frac{L}{2 \delta } ) }{ \frac{L}{2 \delta }} \geq \nu
>0. \]
The result (\ref{eq:4''}) holds to the same order as the energy
expansion (\ref{eq:2}). The filled lowest energy
level \( n=0 \) corresponds with the density \( \nu \) given by:
\begin{equation}
\label{eq:5}
\nu_{1} \frac{tanh( \frac{L}{2 \delta } ) }{ \frac{L}{2 \delta } } = \nu .
\end{equation}
It follows from (\ref{eq:5}) that there is a decrease of the number of
\( n=0 \) states with respect to the uniform field case. In this later case
the density at which the \( n=0 \) state is filled is given by \( \nu_{1}
\equiv \frac{ \Phi}{ \Phi_{0}}. \) Moreover in our limit of large but
still nonhomogeneous fields the quantity \( \nu_{1} \) satisfies the inequality
given above (\ref{eq:2}).

      Let us now compare total energies of our gas at a given density \( \nu \)
between the \( n=0  \) state in the uniform field B and the \( n=0 \)
state in the nonhomogeneous magnetic field B with a finite
parameter \( \delta  \). We obtain from (\ref{eq:1}) and (\ref{eq:4''}) that
their total energy difference is given by:
\begin{equation}
\label{eq:6}
E_{T}(B, \nu ) - E_{T}(B, \delta , \nu ) = N \nu t ( \frac{a}{ \delta })^{2} /2
> 0
\end{equation}
for
\[ \nu_{1} \frac{tanh( \frac{L}{2 \delta })}{ \frac{L}{2 \delta }} \geq \nu > 0. \]
It follows from (\ref{eq:6}) that
in this range of densities and of system parameters values the
nonhomogeneous field state has lower energy than that in the homogeneous field.

          Let us now increase the gas density  \( \nu \) to the value \( \nu_{1}. \)
The lowest energy level of the uniform field state becomes
filled. Let us assume that the nonhomogeneity parameter \( \delta \) is large
enough and such that the following inequalities hold:
\[ \nu_{1} \frac{tanh( \frac{L}{2 \delta })}{ \frac{L}{2 \delta }} < \nu_{1} <
2 \nu_{1} \frac{tanh( \frac{L}{2 \delta })}{ \frac{L}{2 \delta }}. \]
Then the \( n=0 \) level of the nonhomogeneous field case is
filled completely and the \( n=1 \) level of the same case only partially.
Let us compare energies of the uniform field state
and of the nonhomogeneous state for the density  \( \nu_{1} = \nu. \)
We obtain for the difference of the total energies of both states::
\[ E^{n=1}(B, \delta , \nu ) - E^{n=0}( B, \nu ) = \\
4 \pi tN \nu_{1}^{2} (1- \frac{tanh( \frac{L}{2 \delta })}{ \frac{L}{2 \delta }} )
- N \nu_{1} t( \frac{a}{ \delta })^{2}
(5 - 4 \frac{tanh( \frac{L}{2 \delta })}{ \frac{L}{2 \delta }})/2. \]
This quantity is positive for macroscopically nonvanishing
density \( \nu. \) Decrease of the single fermion energy due to the
nonhomogeneity is overcompensated by the decrease of the  number of particles
in the \( n=0 \) nonhomogeneous field level and by their increase
in the \( n=1 \) level. The gap between energies of these two levels is
\[ \hbar \omega - \frac{ \hbar^{2}}{m \delta^{2}} + O(1/F), \]
the increase of the number of particles in the \( n=1 \) level increases
substantially the total energy of the system. Thus the uniform field state
becomes preferred at higher particle densities.
Qualitatively the same type of conclusions holds for higher
densities \( \nu \) and higher level numbers n.

We conclude that the nonhomogeneous field state of our gas of spinless fermions
is preferred with respect to the uniform field state of the same gas
for densities \( \nu \) less or equal to a critical value
\( \nu_{c}(B, \delta) \) defined as
\[ \nu_{c}(B, \delta) \equiv \nu_{1}
\frac{ tanh( \frac{L}{2 \delta } ) }{ \frac{L}{2 \delta } }. \]
For higher densities \( \nu > \nu_{c}(B, \delta ) \) the later state is
preferred.

One may ask at which value of the nonhomogeneity parameter \( \delta \)
the energy difference (\ref{eq:6}) takes the largest value.
The difference \( E_{T}(B, \nu ) - E_{T}(B, \delta , \nu )  \) from
(\ref{eq:6}) becomes larger when \( ( \frac{L}{ \delta })^{2} =
N ( \frac{a}{ \delta } )^{2} \) is increasing quantity, e.i. when
\( \delta \) decreases with respect to the the length L.
There exists a critical value of \( \delta \) given by
\(  \delta_{c} \equiv L/ln( \pi N \nu_{1}), \). It is that limiting value of
\( \delta \) for which the inequality below (\ref{eq:3}) becomes equality.
Below  \( \delta_{c} \) the degeneracy of every energy level becomes
n-dependent \cite{[4]} as it follows from \( p_{y} \) dependence of
\( n_{max} \) given in the beggining of this paper. It is possible to find that
\[ D_{n} \approx D_{L}(2 \delta /L)(1- \frac{2}{F}(n + \frac{1}{2})), \]
if
\[ tanh(L/2 \delta ) > 1- \frac{2}{F}(n + \frac{1}{2} ). \]
The density  \( \nu \) in (\ref{eq:6}) is for the \( n=0 \) state now from
the region:
\[ \nu_{1} (2 \delta /L) (1 - 1/F)  \geq \nu >0. \]
We have found that for such more localized nonhomogeneity of the field
for which \( \delta \) is smaller, \( \delta < \delta_{c}, \) the energy
difference depends on \( ( L/2 \delta ) \) in the same way as in (\ref{eq:6}).
The maximum value of this difference for the filled lowest level
\( n=0 \) of the nonhomogeneous field state is obtained for
\( \delta =a(3/2 \pi \nu_{1})^{1/2} \) as:
\[ E_{T}(B, \nu ) - E_{T}(B, \delta , \nu ) = (4t \sqrt{N}/3) \nu \sqrt{ \pi
\nu_{1}} / 6. \]

The gas density corresponding to this value is found to be
\( \nu =( \sqrt{6}  - \frac{2}{3}) \sqrt{ \nu_{1} / \pi N}.  \)
This density corresponds to
nonzero linear density of particles, \( N_{f}/ \sqrt{N}. \)
Thus we conclude that in our type of the nonhomogeneous field
the energy difference (\ref{eq:6}) is maximized for small densities of
particles. One can say that it is an edge effect which leads to the maximum
of the considered energy difference.

Let us discuss shortly consequences of our results obtained above for
the anyon gas physics. According to \cite{[6]} and \cite{[2']} anyons may be
described as spinless fermions moving in the statistical field generated
by the statistical potential \( {\bf A_{i} } \) acting on the i-th anyon
and given by: \\
\(     {\bf A_{i}} = (1- \rho )( \hbar c/e) {\bf z} \) { \small x} \( \Sigma_{j \neq i}
( {\bf r_{i} - r_{j} } ) / ( \mid r_{i} - r_{j} \mid )^{2}. \) \\ The fractional
statistics parameter is denoted here by  \( \rho . \) A Hartree-Fock
considerations in \cite{[6]} lead to description of anyons with a
single fermion  Hamiltonian
\[ H = \frac{1}{2m} [  { \bf p} - \frac{e}{c} {\bf A} ]^{2}, \]
where the uniform average statistical field is described by the vector
potential \\
\( {\bf A} = (1- \rho )( \hbar c \nu /a^{2}e) {\bf z} \) {\small x}
\( {\bf r}. \)
This potential may be transformed into another gauge form:
\[ {\bf A} = (1- \rho )( \hbar c \nu /a^{2}e)( 0, x-x_{0}, 0). \]
We may assume that the potential of our field
\[ {\bf A} = ( 0, B \delta tanh((x-x_{0})/ \delta ), 0) \]
is a result of the averaging of the potential \( {\bf A}_{i} \) given
above over some stationary configurations of anyons which are
different from those which lead to the uniform statistical 
field. Such a modulated form of the average
potential may results from  \( {\bf A}_{i} \)  when fermions are non-
homogeneously distributed within the plane.

      From our results described above for the gas of spinless fermions
moving in our nonhomogeneous field it follows that for the anyon densities
\( \nu \) such that inequality  \[ \nu_{1} \frac{tanh( \frac{L}
{2 \delta })}{ \frac{L}{2 \delta }} \geq \nu > 0 \]
holds ( where  \(  \nu , L, \delta  \) are fixed parameters ) the uniform
statistical field state is unstable with respect to the state
with nonhomogeneous statistical field described by our 
potential. Our comparison of the total energies between
the nonhomogeneous field state and the uniform field state
as given above shows that the former is preferred at zero temperature.
This instability effect of the uniform statistical field state
is larger for smaller densities. However it 
is present also for macroscopically nonvanishing density of
particles. For densities of anyons \( \nu \) higher than
\( \nu_{1} \frac{tanh( \frac{L}{2 \delta } ) }{ \frac{L}{2 \delta } } \)
the uniform statistical field state is preferred for anyons.
It is however not clear from our results whether our
nonhomogeneous field is the most stable nonhomogeneous field state of anyons
at lower densities of particles. In our calculations we may consider
the characteristic length L to be either the linear dimension of the sample
either it may be a characteristic length of a domain within the sample.
In the later case the whole sample is expected to be covered by similar domains.
From (\ref{eq:6}) it follows that the most preferred value of the
nonhomogeneity parameter \( \delta \) is that value for which
\( \nu = \nu_{c}(B, \delta ) \) when the enrgy difference between both
considered states becomes maximized.

Thus we see that our results are directly related to the physics of anyons.
Experimental evidence for presence of these new physical phenomena
in real materials is controversial nowadays.
There exists some positive evidence for observation of broken T- and/or
P- symmetry in superconductors based on oxidic layers, \cite{[5']}.
Some of these experimental results are interpreted, however, as a negative
evidence or there is no their clear interpretation.
Our results presented here may contribute to better understanding why there exists
variety of different results obtained under different physical conditions
and in different samples in cited above experiments.
If the statistical field varies in the sample then also measured physical
quantities such as the optical axis rotation angle will vary within the
\( CuO_{2} \) plane in cuprate perovskites as it may be found f.e. from the
results of analysis in \cite{[WZ]} of rotation of polarized light reflected from
T- and P- violating phases.
According to results presented in this paper in those samples in which higher
densities of charge carriers occur the uniform statistical field anyon state
may be realized. In those samples where the density of anyons is below some
critical value \( \nu_{c} \) the anyon statistical field becomes spatially
modulated. Whether this modulation is described by the statistical field
considered in our calculations remains an open problem. Here we have shown only
that such a transition exists, from our results we are not able to say which
type of field modulation represents the true ground state of anyons.
We may expect that some other than our modulation may lead to energetically
( we consider T=0 ) more preferred anyon state.

Results of this paper point to principal possibility that
a phase transition between anyon states: uniform statistical field state
and modulated statistical field state occurs when the carrier density is
decreased. Such a phase transition may be experimentally observed under
appropriate conditions. Physical properties of modulated states as well as
their response to external signals probing their nature should be established
in other to improve our understanding of the experimental situation in search
of broken T-/P- symmetries due to presence of anyons. It is known, \cite{[DS]},
that dynamic response of fermions in continuum as well as on the lattice in a
magnetic field may be calculated. This task is beyond the scope of this paper.

\subsection*{Acknowledgement:}
Some of problems considered here O.H. formulated during his
visit in the Theoretical Physics Institute, ETH, 
Zurich. He acknowledges the financial support by the ETH,
which enabled the visit. He would like to express his
gratitude to prof. T.M. Rice, prof. G. Blatter, D. 
Poilblanc, Y. Hasegawa and C. Gross for discussions. O.H. acknowledges also the support by the ICTP, Trieste, especially expresses his sincere thanks to Prof. Yu Lu and Prof. E. Tosatti for
their kind hospitality.


\thebibliography{999}
\bibitem{[0]} Ch. Nayak, S.H. Simon, A. Stern,  M. Freedman, S.D. Sarma, Rev. Mod. Phys. {\bf 80} (2008) 1083
\bibitem{[00]}L. Jiang, G.K. Brennen, A.V. Gorshkov, K. Hammerer, M. Hafezi, E. Demler, M.D. Lukin, P. Zoller, Nature Physics {\bf 4} (2008) 482 - 488 
\bibitem{[01]}Y. P. Zhong, D. Xu, P. Wang, C. Song, et al., Phys. Rev. Lett. {\bf 117} (2016) 110501 
\bibitem{[1]}L.D. Landau and E.M. Lifshitz, Quantum Mechanics,
(Theoretical Physics Course, Vol. III.), Nauka, Moscow, 1974, pp. 522-525, in Russian
\bibitem{[2]}Y. Hasegawa, P. Lederer, T.M. Rice and P.B. Wiegmann, 
Phys. Rev. Lett.  {\bf 63}  (1989) 907-910
\bibitem{[3]}       D. Poilblanc,   Phys. Rev. {\bf B40}  (1989) 7376-7379 \\
                    M. Kohmoto,   Phys. Rev.  {\bf B39} (1989) 11943-11949 \\
                    G. Montambaux,   Phys. Rev. Lett.  {\bf 63}  (1990) 1657 \\
                    Y. Hasegawa, Y. Hatsugai, M. Kohmoto, G. Montambaux,   Phys. Rev.  {\bf B41} (1990) 9174-9182 \\
                    A.A. Belov, Ju.E. Lozovik, V.A. Mandelshtam, JETP Lett. {\bf 51}  (1990) 422-425, in Russian \\
                    A.G. Abanov, D.V. Khveshchenko,   Mod. Phys. Lett.  {\bf 4} (1990) 689-696 \\
                    F. Nori, F. Abrahams, G.T. Zimanyi,   Phys. Rev.  {\bf B41}  (1990) 7277 \\
                    R. Rammal, J. Bellisard,   J. Phys. France {\bf 51} (1990) 2153-2165 \\
\bibitem{[BBR]}   A. Barelli, J. Bellisard, R. Rammal,  J. Phys. France {\bf 51} (1990) 2167-2185 \\
\bibitem{[N]}     B.A. Dubrovin, S.P. Novikov,   Sov. Phys. JETP {\bf 52} (1980) 511 \\
                B.A. Dubrovin, S.P. Novikov,   Sov. Math. Dokl. {\bf 22} (1980) 240 \\
                S.P. Novikov,   Sov. Math. Dokl. {\bf 23} (1981) 298 \\
\bibitem{[6]}       R. B. Laughlin,   Phys. Rev. Lett.  {\bf 60} (1988) 2677 \\
                    C. B. Hanna, R. B. Laughlin, A. L. Fetter,   Phys. Rev. {\bf B40} (1989) 8745-8758 \\
\bibitem{[4]}              O. Hud\'{a}k, Zeitschrift für Physik B Condensed Matter 2 {\bf 88}  239-246 \\
\bibitem{[5']}    K.B. Lyons, J. Kwo, J.F. Dillon Jr.,
G.P. Espinosa, M. McGlashab-Powell, A.P. Ramirez, L.P. Schneemeyer, Phys.Rev.Lett.
{\bf 64} (1990) 2949 \\ H.J. Weber et al.,  Sol. St. Comm. {\bf 76} (1990) 511 \\
 S. Spielman, K. Fesler, C.B. Eom, T.H. Geballe, M.M. Fejer and A. Kapitulnik,
 Phys.Rev.Lett. {\bf 65} (1990) 123-126\\ M.A.M. Gijs, A.M. Gerrits, C.W.J. Beenakker,
Phys.Rev. {\bf B42} (1990) 10789 \\
K.B. Lyons, J.F. Dillon, E.S. Hellman, E.H. Hartford, M. McGlashan-Powell, 
 Phys.Rev. {\bf B43} (1991) 11408 \\
R.F. Kiefl et al. Phys.Rev.Lett. {\bf 64} (1990) 2082 \\
\bibitem{[5]}              R. E. Peierls, Quantum Theory of Solids, Oxford,
Clarendon Press, 1955

\bibitem{[2']} Y-H. Chen, F. Wilczek, E. Witten, B.I. Halperin, Int.J.Mod.Phys.
{\bf B3} (1989) 1001-1067 \\ F. Wilczek, Fractional Statistics and Anyon
Superconductivity, World Scientific, Singapore-New Jersey-London-Hong Kong, 1991

\bibitem{[WZ]}  X.G. Wen, A. Zee,   Phys.Rev. {\bf B43} (1991) 5595
\bibitem{[DS]} B. Doucout, P.C.E. Stamp,   Phys.Rev.Lett. {\bf 66} (1991)2503
\end{document}